\documentclass[12pt]{article}
\renewcommand{\baselinestretch}{1.2}
\usepackage{amsmath}
\usepackage{amssymb}
\usepackage{indentfirst}
\usepackage{amssymb}
\usepackage{amsfonts}
\usepackage{amscd}
\usepackage{amsbsy}
\usepackage{amsthm}
\usepackage{latexsym}
\usepackage{graphicx,color} 
\usepackage[dvipsnames]{xcolor}
\usepackage[colorlinks]{hyperref}
\hypersetup{linkcolor=blue,citecolor=blue,urlcolor=blue}
\usepackage{orcidlink}
\usepackage{hyperref}

\usepackage{titlesec}

\titleformat*{\section}{\large\bfseries}
\titleformat*{\subsection}{\normalsize\bfseries}

\def\nn{\nonumber}       
\def\beq{\begin{eqnarray}}
\def\eeq{\end{eqnarray}}

\def\al{\alpha}
\def\be{\beta}

\def\ga{\gamma}
\def\de{\delta}

\def\ep{\epsilon}

\def\ka{\kappa}
\def\la{\lambda}

\def\pa{\partial}

\def\om{\omega}
\def\ph{\varphi}

\def\th{\theta}
\def\te{\vartheta}

\def\De{\Delta}
\def\La{\Lambda}

\def\Om{\Omega}
\def\Te{\Theta}
\def\Th{\Theta}

\DeclareMathOperator{\cx}{\square}

\begin{document}
\begin{center}
\renewcommand*{\thefootnote}{\fnsymbol{footnote}}

{\large\bf
Conditions for positivity of energy in
\\
superrenormalizable polynomial gravity

}
\vskip 6mm
	
{\bf P\'ublio Rwany B. R. do Vale \orcidlink{0000-0001-9264-2083}}
\footnote{E-mail address: \ publio.vale@gmail.com}

\vskip 6mm

 International Institute of Physics, Universidade Federal do Rio Grande do Norte, 
\\	
59078-970, Natal, Rio Grande do Norte, Brazil

\end{center}

\vskip 10mm
\vskip 2mm

\begin{abstract}

\noindent
 At the quantum level, the polynomial models of gravity with six
and eight derivatives are superrenormalizable, but suffer from higher
derivative ghost and/or tachyonic ghost states. On the other hand,
these models may have advantages in the control of negative effects
of ghosts, compared to the more common fourth-derivative theory.
We explore the positiveness of energy of the individual plane wave
solutions in the general models with six and eight derivatives.
Different from the fourth-order gravity, the part of the energy which may be seen as the leading one in the UV, is positively defined in the tensor sector. We extend this investigation to the scalar sectors of the free theory.
\vskip 3mm


\end{abstract}

\setcounter{footnote}{0} 
\renewcommand*{\thefootnote}{\arabic{footnote}} 

\section{Introduction}
\label{sec1}

The issue of higher derivatives and subsequent physical problems
can be identified as a serious difficulty of quantum gravity (QG). The
renormalizability of the theory requires inclusion of at least
four-derivative terms in the action \cite{UtDW,Stelle77}. Even if
higher derivatives are not introduced at the classical level, the
corresponding terms emerge owing to the loop effects of gravity
\cite{hove,dene}, or quantum matter fields \cite{UtDW,DeWitt2003}.
Thus, the theory without at least four derivatives cannot be
renormalizable. This means, the theory should have negative-energy
states, or higher derivative ghosts.
Historically, higher-derivative instabilities were described in
classical mechanics by M.V.~Ostrogradky more than one and
a half  centuries ago \cite{Ostrog} (see also recent discussion in
\cite{Woodard-review}). At the quantum level, the presence
of a ghost generates instabilities usually characterized as the
violation of unitarity \cite{Veltman-63,Stelle77}.  This situation
was extensively discussed in the fourth-derivative quantum gravity,
but can be extended to more general versions of the theory.

There were several interesting proposals concerning
the contradiction between renormalizability and unitarity
in QG. We can invoke the pioneer works
\cite{Tomboulis-77,salstr,antomb}, describing the situations when
the massive ghosts in fourth-derivative gravity become unstable
and eventually decay at the quantum level, such that the unitarity
of the $S$-matrix is restored. The question
of whether this really happens or not, remains open because of the
limited information about higher loop corrections in QG
\cite{Johnston}.

The ``minimal'' renormalizable QG theory can be generalized by
introducing more derivatives. One of the possibilities is a
polynomial QG with six or more derivatives \cite{highderi}, that
gives superrenormalizable models. In this case, the divergences
show up only at the first loop. Furthermore, only four-, second-
and zero-derivative terms in the action get renormalized. Starting
from ten derivatives, all divergences are at the one-loop level. The
same situation is typical for the non-polynomial QG models
\cite{Krasnikov,Kuzmin,Tomboulis-97}, as can be seen from the
corresponding power counting \cite{CountGhosts}. In the polynomial
superrenormalizable QG, all massive poles may be complex and
the corresponding quantum theories unitary
within the Lee-Wick quantization scheme \cite{Modesto-complex}.
Thus, what requires a lot of information about dressed propagator
of gravitons in the four-derivative QG \cite{salstr,Johnston}, may
be automatic in the superrenormalizable models. An unfortunate
detail is that the unitarity of the $S$-matrix does not guarantee the
physical unitarity \cite{Asorey2018} or the stability of classical
solutions.

A radical solution to the problem of ghosts has been formulated in
the proposal of Simon \cite{Simon-90,ParSim}. In this approach,
higher-derivative terms are treated, by definition, as small
perturbations over the Einstein-Hilbert action. In this scheme,
there are no ghosts, the stability of solutions and the unitarity
of the quantum theory are guaranteed. However, this alternative
construction of QG crucially differs from the standard construction
of usual quantum field theory. On top of that, it generates a large
portion of ambiguities \cite{Maroto1997} (see also \cite{Simon-91}
and the discussion in \cite{ABS}). Both problems may be solved
if, instead of defining higher derivatives as small perturbations, we
impose the restrictions on the initial conditions, such that the ghost
cannot be generated from the vacuum
\cite{HD-Stab,PP,Salvio-19,LinNonlin}. Creating
a ghost from nothing requires the concentration of gravitons with
the Planck energy density, $M_P^4 \approx 10^{76}$ GeV$^4$.
The unclear aspect in this area is to explain why there is such a
restriction on the energy density of gravitons.

Another idea to fight ghosts, in a completely different direction,
was proposed by Hawking in \cite{Hawking-Whois}. The point is that,
in the fourth-order theory, both the graviton and the massive ghost
are not independent particles, but  parts of the same quantum
excitation.
E.g., the propagator of metric perturbations can be separated into
two or more parts, but originally these parts belong to the same
combined quantum state. It might happen that taking this aspect into
account, one achieves stability. The stable combinations of complex
ghost-like states were called ghostballs \cite{LiuModestoCalcagni},
but in may also include usual ghosts and healthy degrees of freedom.
The consistent description of a ghost confinement may be an ultimate
solution of the problem of higher derivatives. At the moment, such a
solution is not elaborated in detail. Such elaboration requires, in
particular, a better understanding of different aspects of the theory
with higher than four derivatives. In the present work, we
concentrate on the particular issue of energy positivity in these
models, which may be relevant for better general understanding of
the superrenormalizable theories.

An important aspect of superrenormalizable QG models is the
definition of signs in the action, which makes the
massless graviton to be a healthy particle and provides the mass
spectrum of the regular form, as described in \cite{highderi}. In
the present work, we explore this issue using the simple method
applied recently in the book \cite{OUP} for the sign definition in
the fourth derivative QG. As we will see in what follows, in the
UV (high frequency) regime, the energy of the combination of
tensor ghosts and normal degrees of freedom in the free sector of
six-derivative models may be positively defined at leading order in this energy scale, while in the
eight-derivative model it is negatively defined. The sign rules
can be established for both gauge-fixing independent, tensor and
scalar, sectors of the propagator.

The paper is organized as follows. Sec.~\ref{sec2} reviews the
propagator of a general model of QG with an arbitrary number of
derivatives. Here we closely follow the book \cite{OUP} and the
review paper \cite{QG-review} and this section is included to make
our work self-consistent. We identify the gauge-independent
tensor and scalar sectors of the theory. Sections \ref{sec3}
and \ref{sec4} report about the main result, that is the conditions
of positivity of the plane wave solutions for tensor and scalar and
modes in the theory with six and eight derivatives. Cumbersome
technical details are separated in the Appendix. Sec.~\ref{sec5}
describes the situation in the theories with more than eight
derivatives. Finally, in Sec.~\ref{sec6}, we draw our conclusions.

\section{Propagator in the higher-derivative gravity}
\label{sec2}

In the present work, we consider the energy conditions for the
free field limit of the higher-derivative QG of a general
form.\footnote{A more detailed version of this section can be
found in the Handbook on QG \cite{QG-review}.}
The parametrization of the metric on the flat background is given by
\beq
\label{hmn}
g_{\mu\nu} \,=\, \eta_{\mu\nu} + h_{\mu\nu} \,.
\eeq
Since the curvature tensor is $\mathcal{O}(h_{\mu\nu})$, the free
limit means we need the terms in the action which are at most of the
second order in curvatures. Higher order terms and the total
derivative terms in the Lagrangian may be omitted. In this framework,
the most general action of the higher-derivative gravity is
\beq
\label{eq:1.1}
S  
\,=\,
\frac{1}{2}
\int d^{4}x\sqrt{-g}\,\Big\{
C_{\mu\nu\al\be} \Phi\left(-\square\right)C^{\mu\nu\al\be}
+ R\Psi \left(-\square\right)R
- \frac{2}{\ka^2}\, R\Big\} \,,
\eeq
where the cosmological constant term is assumed zero, because we
need to use the flat background, $C_{\mu\nu\al\be}$ is the Weyl
tensor, $R$ is the scalar curvature, $\Phi (-\square)$ and $\Psi (-\square)$ are
functions which are chosen such that the QG theory is
superrenormalizable. For instance, $\Phi$ and $\Psi$ may be
polynomials of the same order. Similar conditions can be also
formulated in nonlocal higher-derivative models
\cite{Modesto-nonloc,OUP}. In the present section, we do not
assume any restrictions on these functions, but subsequently
concentrate on the polynomial cases only.

One can rewrite the Lagrangian in (\ref{eq:1.1}) in another
basis, using the relations
\beq
&&
C_{\mu\nu\al\be}\Phi\left(-\cx\right)C^{\mu\nu\al\be}
\,=\,
R_{\mu\nu\al\be}\Phi\left(-\cx\right)R^{\mu\nu\al\be}
- 2 R_{\mu\nu}\Phi\left(-\cx\right)R^{\mu\nu}
+ \dfrac{1}{3}R\Phi\left(-\cx\right)R\,,
\qquad
\label{eq:1.2}
\\
&&
\textrm{GB}_{n}
\,\equiv\,
R_{\mu\nu\al\be}\Om\left(-\cx\right)R^{\mu\nu\al\be}
- 4R_{\mu\nu}\Om\left(-\cx\right)R^{\mu\nu}
+ R\Om\left(-\cx\right)R
\,=\, \mathcal{O}\left(R_{\cdots}^{3}\right) \,,
\label{eq:2.6}
\eeq
where $\Om\left( -\square \right)$ may be polynomial of degree $n$. The
Gauss--Bonnet-like term
$\textrm{GB}_{n}$ is certainly not topological for $n\neq0$,
but it contributes only to the third- and higher-order terms in the
curvature tensor. The relation \eqref{eq:2.6} is proved in
\cite{QG-review,BreTibLiv,Tib-notes}, including in the case
of non-polynomial functions.

To construct the propagator, we introduce the gauge-fixing term
\beq
\label{1.11}
S_{\textrm{gf}}
\,=\,
\dfrac{1}{2}
\int d^{4}x\,\,\chi_{\mu}\,\mathcal{Y}^{\mu\nu} \chi_{\nu}\,,
\eeq
where the gauge condition depends on the parameter $\be$,
\beq
\label{1.12}
\chi_{\mu} \,=\, \pa_{\la}h^\la_{\,\,\mu}
- \be\pa_{\mu}h\,,
\eeq
being $h = \eta^{\mu\nu} h_{\mu\nu}$ and the weight
operator includes the form factor $\mathcal{W}(-\cx)$,
\beq
\label{1.13}
\mathcal{Y}^{\mu\nu}
\,=\,
\big(\eta^{\mu\nu}\cx \,-\, \ga\,\pa^{\mu}\pa^{\nu}\big)
\mathcal{W}(-\cx )\,.
\eeq
The ambiguities include the gauge fixing parameters  $\be$,
$\ga$ and the operator function $\mathcal{W}(-\cx )$. All these
quantities should be chosen to remove the degeneracy of the total
bilinear form, but this leaves an arbitrariness.

The propagator depends on the quadratic expansion of the
action, i.e., of the expression
\beq
\big( S + S_{\textrm{gf}}\big)^{(2)}
&=&
\int d^{4}x \,h^{\mu\nu}\,H_{\mu\nu,\al\be}\,
h^{\al\be} \,.
\label{eq:666}
\eeq
Omitting the technical details (which can be found in \cite{OUP}
or, in a more complete form, in \cite{QG-review}), let us start with
the expression for the degenerate bilinear form of the action with
the gauge fixing term, in the momentum representation,
\beq
&&
\hat{H}
\,=\,
H_{\mu\nu,\al\be}(k)
\,=\,
s_{1}(k^2)\de_{\mu\nu,\al\be}
+ s_{2}(k^2)\eta_{\mu\nu}\eta_{\al\be}
+ s_{3}(k^2)\left(\eta_{\mu\nu}k_{\al}k_{\be}
+\eta_{\al\be}k_{\mu}k_{\nu}\right)
\nn
\\
&&
\qquad
+\,\,
s_{4}(k^2)\left(\eta_{\nu\be}k_{\mu}k_{\al}
+\eta_{\mu\be}k_{\nu}k_{\al}
+\eta_{\nu\al}k_{\mu}k_{\be}+\eta_{\mu\al}k_{\nu}k_{\be}\right)
\,+\, 
s_{5}(k^2)k_{\mu}k_{\nu}k_{\al}k_{\be} \,,
\label{1.9}
\eeq
with the coefficient functions
\beq
\label{1.17}
s_{1}(k^2) & = & \dfrac{1}{2}\Phi(k^2) k^4
+\dfrac{1}{2\ka^2}\,k^2 \,,
\nn
\\
s_{2}(k^2)
& = & \Big[-\dfrac{1}{6}\Phi(k^2)+\Psi(k^2) \Big]k^4
- \dfrac{1}{2\ka^2}k^2
+ \be^2 (\ga-1)\mathcal{W}(k^2)k^4 \,,
\nn
\\
s_{3}(k^2)
& = &
\Big[\dfrac{1}{6}\Phi(k^2)-\Psi(k^2) \Big] k^4
+ \dfrac{1}{2\ka^2}k^2 - \be (\ga-1)\mathcal{W}(k^2)k^4 \,,
\nn
\\
s_{4}(k^2)
& = & - \,\,\Phi(k^2) k^4 - \dfrac{1}{\ka^{2}}k^2
- \mathcal{W}(k^2)k^4 \,,
\nn
\\
s_{5}(k^2)
& = &
\Big[\dfrac{1}{3}\Phi(k^2)+\Psi(k^2)\Big]k^4
+\ga\mathcal{W}(k^2)k^{4}\,.
\eeq
The bilinear form of the total action (i.e., the inverse of the 
propagator) depends on $\Phi(k^2)$, $\Psi(k^2)$, on the gauge-fixing parameters, including the function $\mathcal{W}(k^2)$. One can
check that the coefficient of the ``generalized'' Gauss-Bonnet term
$\Om(k^2)$ does not appear in these expressions.
It is also noteworthy that $s_1(k^2)$ is the only coefficient that does not depend on the terms arising from the gauge-fixing action.

It is useful to rewrite the operator
$H_{\mu\nu,\al\be}$ using Barnes-Rivers
projectors, defined in terms of the longitudinal
$\,\om_{\mu\nu} = k_\mu k_\nu/k^2\,$ and transverse
$\,\th_{\mu\nu} = \eta_{\mu\nu} - \om_{\mu\nu}$
projectors in the vector space. The tensor projectors are
\beq
{\hat P}^{(2)}
&=&
P^{(2)}_{\mu\nu\,,\,\al\be}
\,=\,
\frac12(\theta_{\mu\al}\theta_{\nu\be}
+ \theta_{\mu\be}\theta_{\nu\al})
- \frac{1}{3}\,\theta_{\mu\nu}\theta_{\al\be}\,,
\nonumber
\\
{ \hat P}^{(1)}
&=&
P^{(1)}_{\mu\nu\,,\,\al\be}
\,=\,
\frac12\,(\theta_{\mu\al}\om_{\nu\be} + \theta_{\nu\al}\om_{\mu\be}
+ \theta_{\mu\be}\om_{\nu\al} + \theta_{\nu\be}\om_{\mu\al})\,,
\nonumber
\\
{\hat P}^{(0-s)}
&=&
P^{{(0-s)}}_{\mu\nu\,,\,\al\be}
\,=\,
\frac{1}{3}\,\theta_{\mu\nu}\theta_{\al\be} \, ,
\quad
{\hat P}^{(0-w)}
\,=\,
P^{{(0-w)}}_{\mu\nu\,,\,\al\be}
\,=\,
\om_{\mu\nu}\om_{\al\be} \, .
\mbox{\qquad}
\label{proj-tensor} 
\eeq
To complete the algebra, we need also the transfer operators
\beq
{\hat P}^{{(ws)}}
= P^{(ws)}_{\mu\nu\,,\,\al\be}
= \frac{1}{\sqrt{3}}\,\theta_{\mu\nu}\om_{\al\be}
\, ,\quad
{\hat P}^{{(sw)}}
= P^{(sw)}_{\mu\nu\,,\,\al\be}
= \frac{1}{\sqrt{3}}\,\om_{\mu\nu}\theta_{\al\be} \, .
\mbox{\qquad}
\label{projBR}
\eeq
In this basis, (\ref{1.9}) is cast in the form
\beq
\hat{H}
\,=\,
b_{2}\hat{P}^{(2)} + b_{1}\hat{P}^{(1)}
+ b_{os}\hat{P}^{(0-s)} + b_{ow}\hat{P}^{(0-w)}
+ b_{sw}\big[\hat{P}^{\left(ws\right)}
+ \hat{P}^{(sw)}\big] \, ,
\label{1.18}
\eeq
where
\beq
\label{1.20}
b_{2}&=&\dfrac{1}{2}k^{4}\Phi(k^2)+\dfrac{k^2}{2\ka^2} \, ,
\nn
\\
b_{1}&=&-\dfrac{k^{4}}{2}\mathcal{W}(k^2) \, ,
\nn
\\
b_{os}&=&-\dfrac{k^2}{\ka^2} + 3k^4\left[\Psi(k^2)
+ \be^2\left(\ga-1\right)\mathcal{W}(k^2)\right] \, ,
\nn
\\
b_{ow}&=& k^{4}\left(\be-1\right)^2
(\ga-1)\mathcal{W}(k^2) \, , 
\nn
\\
b_{sw}&=&\sqrt{3}\left[k^{4}\be\left(\be-1\right)
(\ga-1)\mathcal{W}(k^2)\right] \, .
\eeq
Finally, using these relations, we arrive at the propagator
\beq
\label{1.21}
G(k)
&=&
\dfrac{2\ka^2}{k^{2}\left[\ka^2\Phi (k^{2})k^2+1\right]}\,
\hat{P}^{(2)}
- \dfrac{2}{\mathcal{W}(k^2)k^{4}}\,\hat{P}^{(1)}
+ \dfrac{\ka^2}{k^2\left[3\ka^2\Psi(k^2)k^2-1\right]}\,\hat{P}^{(0-s)}
\nn
\\
&&
+\,\,\dfrac{1}{\left(\be-1\right)^{2}}
\bigg[\dfrac{3\be^{2}\ka^2}{k^{2}
	\left[3\ka^2\Psi(k^2)k^2 - 1\right]}
+ \dfrac{1}{\mathcal{W}(k^2)(\ga-1)k^{4}}\bigg]\hat{P}^{(0-w)}
\nn
\\
&&
- \,\,
\dfrac{\sqrt{3}\be\ka^2}{\left(\be-1\right)k^2
	\left[3\ka^2\Psi(k^2)k^2-1\right]}
\big[\hat{P}^{(ws)}+\hat{P}^{(sw)}\big] \, .
\eeq
The form factor $\Phi (k^{2})$ of the Weyl-squared term completely
defines the propagation of the spin-2 mode.
The scalar mode with $\hat{P}^{(0-s)}$ depends only on the form
factor $\Psi (k^{2})$ of the $R^2$-term. These two
parts do not depend on the gauge fixing, while all other sectors are
gauge dependent. An interesting detail is that the
gauge-fixing independence of the  $\hat{P}^{(0-s)}$-mode
holds only in four spacetime dimensions and for $\La=0$
\cite{QG-review}.

An alternative representation of the metric perturbations (\ref{hmn}) can be obtained by decomposing them into their spin-2, spin-1, and spin-0 modes, namely,
\beq
\label{2.2}
h_{\mu\nu}
\,=\,
\bar{h}_{\mu\nu}^{\perp\perp}
+ \pa_{\mu}\epsilon_{\nu}^{\perp}
+ \pa_{\nu}\epsilon_{\mu}^{\perp}
+ \pa_{\mu} \pa_{\nu} \epsilon
+ \phi_{\mu\nu} \, .
\eeq
Here the spin-2 mode $\bar{h}_{\mu\nu}^{\perp\perp}$ is traceless
and transverse, i.e.,
\beq
\label{2.3}
\bar{h}_{\mu\nu}^{\perp\perp}\eta^{\mu\nu}
\,=\,
0\quad\textrm{and}\quad\pa^{\mu}\bar{h}_{\mu\nu}^{\perp\perp}=0
\, ,
\eeq
while spin-1 part satisfies $\pa_{\mu}\epsilon^{\perp \mu}=0$.
The two scalar modes $\ep$ and $h$ can be combined into the
gauge-fixing independent field
\beq
\label{2.5}
\phi_{\mu\nu}\,=\,\frac{1}{3} \theta_{\mu\nu} ( h - \cx\ep )  \, ,
\qquad
\mbox{such that}
\qquad
\hat{P}^{(0-s)}_{\mu\nu,\al\be}\,\phi^{\mu\nu}
\,\,=\,\,\phi_{\al\be} \, .
\eeq
The bilinear action with gauge-fixing independent terms can be written as
\beq
\label{2.4}
S^{(2)} \,=\,  \dfrac{1}{2}\int d^4x\left[\bar{h}^{\perp\perp\mu\nu}
H_2(-\cx)\,\bar{h}_{\mu\nu}^{\perp\perp}
\,+\,
\phi^{\mu\nu}H_{0}\left(-\square\right)\phi_{\mu\nu}\right] 
\, ,
\eeq
where the operators $H_2(-\cx)$ and $H_0(-\cx)$ can be read off
from \eqref{1.21}, namely,
\beq
\label{1.23}
&&
H_{2}(-\cx)\,=\,\dfrac{\cx}{2\ka^2}\left[\ka^2\Phi(-\cx )\cx-1\right] \, ,
\nn
\\
&&
H_{0}(-\cx)\,=\,\dfrac{\cx}{\ka^2}\left[3\ka^2\Psi(-\cx )\cx+1\right] \, .
\eeq

The spectrum of the model (\ref{eq:1.1}) depends of the position of
the poles of the gauge-independent part of the propagator, that means
the roots of the equations
\beq
\label{eqroots}
\ka^2\Phi (k^{2})k^2+1 = 0
\qquad \text{and} \qquad
3\ka^2\Psi(k^2)k^2-1 = 0 \, ,
\eeq
for the massive modes of spin-2 and of spin-0, respectively.
If both $\Phi$ and $\Psi$ are non-zero constants, the theory
is the fourth-derivative gravity. Then, besides the massless
graviton, there are a massive scalar and a massive tensor mode.
If  $\Phi$ and $\Psi$ are polynomials of degree $N$,
then each of the equations~\eqref{eqroots} can have up to $N+1$
distinct roots in the complex plane, that gives $N+1$ massive
modes of spin-2 and of spin-0.

The nature of massive modes depends on the polynomials $\Phi$ and
$\Psi$. In the recent years, there has been an increasing interest in
the models with complex poles (see e.g., \cite{AKPShap (25)}), owing to the possibility of getting
the theory with unitary $S$-matrix in the Lee-Wick quantum
gravity~\cite{Modesto-complex,Modesto16}. Further applications can
be found, e.g., in~\cite{ABS,BreTib1,NosLW} and references therein.
Higher-derivative gravity models with degenerate poles were studied
also in~\cite{ABS,BreTib1}, however in the present work we shall
not consider this possibility.

\section{Positivity of energy in six-order gravity}
\label{sec3}

It is known that in the fourth-derivative gravity the conditions for
positive energy  for individual plane waves are helpful in defining
the signs of the Weyl-squared and $R$-squared terms \cite{OUP}.
Adjusting the coefficients of these terms in the action, one can
provide that:

\textit{i)} The IR limit, i.e., the general
relativity (GR), has positively defined spin-2 sector;

\textit{ii)}  The energy of the massless tensor mode (graviton) is
positive at any energy scale;

\textit{iii)} The sign of the overall energy in the low-energy (IR)
limit is positive.

The first condition \textit{i)} is the cornerstone of the effective
approach to gravity, which assumes that GR is the universal
low-energy limit of any theory of quantum gravity \cite{don}.
Recently there was an indirect convincing confirmation of this
assumption  \cite{EffAM} and here we consider this as a proved
aspect of quantum gravity.
The last condition \textit{iii)} does not depend on higher-order
time derivatives and is relevant for a consistent application of the
effective approach in the IR. The constraints on the signs of the
coefficients in the action, obtained from these conditions, confirmed
the previous well-known results of  \cite{stelle78}.

Let us apply the same method 
to the superrenormalizable polynomial QG,
\beq
S_{2N+4}   &=&
-\dfrac{1}{2}\int \!d^4x \sqrt{-g}\, \Big\{
 \te_{N,C\,} C \, \square^{N} C
+ \te_{N,R} R \, \square^{N} R
+ \te_{N,\mathrm{GB}}\mathrm{GB}_{N}
\nonumber
\\
&& 
+\,\te_{N-1,C\,} C \, \square^{N-1}C
+  \te_{N-1,R}R \, \square^{N-1}R
+ \te_{N-1,\mathrm{GB}}\mathrm{GB}_{N-1}
\nonumber
\\
&& 
+ \,\,\, \ldots \,\,\, +
\nonumber
\\
&& 
+\, \te_{0,C\,}C^2
+ \te_{0,R}R^2
+ \te_{0,\mathrm{GB}}\mathrm{GB}_{0}
+ \dfrac{2}{\ka^2}\,
+ {\mathcal O}(R_{\dots}^3)
\Big\}\,.
\mbox{\qquad}
\label{gaction}
\eeq
Here  $\,C \cx^N C\,$ is the Weyl-squared term in (\ref{eq:1.1})
with $\Phi\left(-\cx\right) = \cx^N$ and  $\mathrm{GB}_N$ are
the ``generalized'' Gauss-Bonnet terms (\ref{eq:2.6}) with
$\Om(-\cx) = \cx^N$. As we already know, the latter terms reduce
to ${\mathcal O}(R_{\dots}^3)$-terms and may be omitted.
In this section, we consider the simplest case of $N=1$, so the
action of our interest is
\beq
S_6
\,=\,
-\,\dfrac{1}{2}\int \!d^4x \sqrt{-g}\,
\Big\{
\te_{1,C\,} C  \cx  C
+ \te_{1,R} R \cx R
+ \te_{0,C\,}C^{2} + \te_{0,R}R^2 
+ \dfrac{2}{\ka^2}\,R\Big\} \, .
\mbox{\qquad}
\label{6action}
\eeq
As in the case of general polynomial action (\ref{gaction}), all
$\te$'s are arbitrary parameters, which values can be defined only
from measurements (experiments or observations).
From the theoretical side, any values of these parameters
do not contradict superrenormalizability, and can be chosen
to avoid problems with ghosts and tachyonic ghosts.

The analysis of signs in \cite{OUP} concerned the case of
$\te_{1,R} = \te_{1,C}=0$,  and dealt with the tensor sector only.
The idea was to consider an individual wave with the space 
wave vector $\vec{k}$ and with a certain frequency. Requiring the
positiveness of the energy of graviton (massless mode) at the
different energy scales, one can show that  $\te_{0,C}  > 0$.

\subsection{Tensor sector in the six-order gravity}
\label{sec3.1}

For the spin-2 part of the action of a six-derivative gravitational model,
we use the first term on the right-hand side (\textit{r.h.s.}) of Eq.
(\ref{2.4}) with the form factor
\beq
\Phi(-\cx ) = -\,\te_{1,C}\cx - \frac{1}{\la}
\,.
\label{lambdaGCC}
\eeq
We redefined the coefficient $\te_{0,C} \equiv 1/\la$, following the
notation used, e.g., in~\cite{OUP}. In the pure tensor sector, the
indices are irrelevant and we can denote, for the sake of brevity,
$\bar{h}_{\mu\nu}^{\perp\perp} = h$. Therefore, the bilinear
form reads
\beq
\label{3.2}
{S}_{\textrm{spin-2}}^{(2)}
\, = \,-\,\,\dfrac{1}{4}\int d^4 x
\Big\{ \te_{1,C} \left(\cx^{2}h\right)\cx h
\,+\, \frac{1}{\la} \left(\cx h\right)^{2}
\,+\,  \frac{1}{\ka^{2}}\,h\cx h
\Big\} \, .
\eeq

The Lagrange function of the individual wave with the fixed
wave vector $\vec{k}$ can be easily obtained by the replacement
(overdot means a time derivative)
\beq
\label{3.3}
\cx h
\,=\,
\ddot{h} - \De h \, ,
\eeq
after applying the Fourier transform 
\beq
\cx h
\,\,\, \xrightarrow \,\,\,
\ddot{h}+ k^2 h
\quad
\textrm{with}
\quad
k^2 = \vec{k}\cdot \vec{k} \, ,
\eeq 
in a similar manner
\beq
\label{3.4}
\cx^2h
\,=\,
\left(\pa_{t}^{2} - \De\right)^2 h
\,=\,
\left(\pa_{t}^{4} - 2\De\pa_{t}^{2}+\De^{2}\right)h
\,\,\, \xrightarrow \,\,\,
h^{\textrm{(IV)}} + 2 k^{2}\ddot{h}
+ k^4 h \, .
\eeq
After a small algebra, the Lagrange function with fixed
$\vec{k}$ is obtained in the form
\beq
\label{3.5}
L_{\textrm{spin-2}}
&=&
-\,\, \dfrac{1}{4}\,\te_{1,C}\big[
h^{\textrm{(IV)}} + 2 k^{2}\ddot{h}
+ k^{4}h\big] \big(\ddot{h}+k^{2}h\big)
\nn \\ &&
\quad
-\,\,
\dfrac{1}{4\la}\big(\ddot{h}+{k}^{2}h\big)^2
- \dfrac{1}{4 \ka^2}\, h\big(\ddot{h}+{k}^{2}h\big) \, .
\eeq

The equation of motion for tensor perturbations can be derived
from (\ref{3.5}) using the generalized Euler-Lagrange equation  in function of the coordinates $q$, which depends on up to $n$th-order time derivatives, for 
$L=L\left(q,\dot{q},\,\ldots \,,q^{(n)}\right)$
\beq
\dfrac{\pa L}{\pa q}
+\sum^{n}_{j=1}\left(-1\right)^{j}
\dfrac{d^{j}}{d t^{j}}\dfrac{\pa L}{\pa q^{(j)}}=0 \, .
\eeq
In this case, for $n=4$ we arrive at the equation for the individual wave
\beq
\label{3.7}
&&
h^{\textrm{(VI)}}+3{k}^2h^{\textrm{(IV)}}
+ 3{k}^4\ddot{h}+{k}^6h
+ \dfrac{1}{\la\,\te_{1,C}}\big(h^{\textrm{(IV)}}
\,+\, 2{k}^{2}\ddot{h}+{k}^4h\big)
\nn
\\
&&
\qquad \qquad
+ \,\,\dfrac{1}{\ka^2 \te_{1,C}}\big(\ddot{h}
+ {k}^2h\big) \,=\, 0 \, .
\eeq
Eq. (\ref{3.7}) can be recast in the form of a harmonic oscillator–like equation
\beq
\label{3.9}
\Big(\dfrac{\pa^2}{\pa t^2}+{k}^{2}\Big)
\Big(\dfrac{\pa^2}{\pa t^2}+\om^{2}_{1}\Big)
\Big(\dfrac{\pa^2}{\pa t^2}+\om^{2}_{2}\Big)h\,
= \, 0 \,,
\eeq
where
\beq
\om^2_1 = {k}^{2}+m^{2}_{1}
\qquad
\mbox{and}
\qquad
\om^2_2={k}^{2}+m^{2}_{2} \,.
\label{albe}
\eeq
The two massive parameters satisfy the conditions
\beq
\label{3.8}
m^2_1+m^2_2  
\,=\,\dfrac{1}{\la\,\te_{1,C}}
\qquad
\mbox{and} \qquad
m^2_1\, m^2_2
\,=\,\dfrac{1}{\ka^2\,\te_{1,C}} \, .
\eeq
The solution of these systems of equations leads to
\beq
m_1^2
\,=\,
\dfrac{1}{2\la \,\te_{1,C}}
\left(\, 1\,\pm\,
\sqrt{1 - \frac{4\la^2 \te_{1,C}}{\ka^2}}\,\right) \, ,
\nn
\\
m_2^2
\,=\,
\dfrac{1}{2\la \,\te_{1,C}}
\left(\, 1\,\mp\,
\sqrt{1 - \dfrac{4\la^2 \te_{1,C}}{\ka^2}}\,\right) \, .
\label{3.10}
\eeq
Each of the quantities $m_1^2$ and $m_2^2$ can be either real
(positive or negative) or form a complex conjugate pair.

For the plane wave, there may be either oscillator or
anti-oscillator type solutions. For a spectrum consisting only
of normal particles, with positive squares of the masses, there
are only oscillatory solutions to the linearized field equation
(\ref{3.9}),
\beq
h(t)
\,&=&\,
C_0 \cos (k\, t + \ph_0)
+ C_1 \cos (\om_1 t + \ph_1)
+ C_2 \cos (\om_2 t + \ph_2) \, ,
\label{oscila}
\eeq
with $\om$ being real frequencies, arbitrary amplitudes  $C_{0,1,2}$  and
irrelevant phases $\ph_{0,1,2}$.

It is clear that the mass spectrum defined by (\ref{albe}) and
(\ref{3.10}) admits both normal oscillatory solutions (\ref{oscila}),
or the tachyonic-type
exponential solutions. We need $\ka^2 > 0$ as a condition
of stability in the deep IR region, where all massive modes
become irrelevant. On the other hand, this logic does not
apply to $\la$.
We know that $\la > 0$ is a necessary condition of stability
in the low-energy domain in fourth-derivative gravity
\cite{Stelle77,stelle78}.
However, in six-derivative models, there is, typically, no strong
hierarchy of the masses linked to fourth-derivative and
six-derivative terms \cite{ABS2}. Thus, we have to formulate the
conditions for the independent free parameters $\la$ and $\te_{1,C}$
without using the fourth-derivative consistency condition.

Using the definitions  (\ref{albe}) and expressions (\ref{3.10}),
one can easily arrive at the solution to this problem, which requires,
simultaneously,
\beq
\la > 0 \, ,
\qquad
\te_{1,C}>0
\qquad
\mbox{and}
\qquad
\dfrac{4\la^2 \te_{1,C}}{\ka^2} < 1 \, .
\label{3conds}
\eeq
Satisfying these conditions, there are no tachyonic solutions
for any magnitude of $\vec k$. Let us
note that even with $m_1^2$ or $m_2^2$ being negative,
there are also oscillatory solutions (\ref{oscila}), but only for
a sufficiently large absolute value of wave vector $\vec k$.
If the last of the inequalities (\ref{3conds}) is violated, then the
quantities $m_1^2$ and $m_2^2$  form a complex-conjugate pair.
In this case, one must have $\te_{1,C}>0$, while there is no
restriction on the sign of $\la$.

The conditions derived above coincide with the analysis of the
poles of the spin-2 part of the propagator~\cite{ABS}. Indeed,
according to the discussion involving Eq.~\eqref{eqroots}, this
model contains two massive modes of spin-2, that are the roots
$z=\mu^2_{\pm}$ of the equation
\beq
\Big( \te_{1,C} \, z - \frac{1}{\la}\Big)\ka^2  \,  z+1 = 0\, ,
\eeq
namely,
\beq
\label{mus}
\mu^2_{\pm} =
\dfrac{1}{2\la \,\te_{1,C}}
\bigg(\, 1\,\pm\, \sqrt{1 - \frac{4\la^2 \te_{1,C}}{\ka^2}}\,\bigg)\,.
\eeq
In this way, we have the correspondence
\beq
m_1^2 = \mu^2_{\pm}
\quad \text{ and } \quad 
m_2^2 = \mu^2_{\mp}\,.
\eeq

The energy of the theory with
$L=L\left(q,\dot{q},\,\ldots \,,q^{(n)}\right)$,
can be derived using the formula
\beq
\label{energy}
E \,=\,
\sum_{l=1}^{n}\sum_{j=1}^{l}\left(-1\right)^{l+j} q^{(j)}
\dfrac{d^{\,l-j}}{dt^{\,l-j}}\dfrac{\pa L}{\pa {q^{(l)}} } - L \, ,
\eeq
where the above equation determines the energy related  to the Lagrangian
which is a function of the coordinates $q$ and their time derivatives
up to the order $n$.
Let us consider the energy conditions in the six-derivative gravity theory. Taking $n=4$, the energy becomes
in the case of the Lagrange function (\ref{3.5}), 
\beq
\label{3.17}
E^{(6)}_{\textrm{spin-2}}
&=&
\dfrac{1}{4} \Big\{
\te_{1,C} \big(  \dddot{h}^{\,2}
- 2h^{\textrm{(IV)}}\ddot{h}
+ 2h^{\textrm{(V)}}\dot{h}\big)
+ \Big( \dfrac{1}{\la} + 3 \te_{1,C}{k}^{2} \Big)
\big(2\dddot{h} \dot{h} - \ddot{h}^2\big)
\nn
\\
&&
+ \,\,\,
\Big(\dfrac{1}{\ka^{2}}
+\dfrac{2}{\la}{k}^2
+3\te_{1,C} k^4
\Big)\,\dot{h}^{2}
\,+\,
\Big( \dfrac{1}{\ka^{2}}{k}^2
+ \dfrac{1}{\la}{k}^4
+ \te_{1,C}{k}^6 \Big) \,h^2\Big\}\, ,
\eeq
this formula gives
the energy of the individual wave with the wave vector $\vec{k}$.
The expression above enables drawing conclusions, which
we organize in the following order:

	\textit{i)} \ \
	The second and subsequent terms  on the \textit {r.h.s.} in
(\ref{3.17}) have fourth and
	lower orders in time derivatives, while the first has the highest,
	sixth-derivative terms. However, part of the terms with lower-order
	time derivatives are coming from the six-derivative terms in the
	action, as they include positive powers of $k^2$;
	
	\textit{ii)} \ \
	Taking $\te_{1,C}=0$,
	we obtain the energy of a fourth-derivative gravity model,
	as it was found in \cite{OUP},
	\beq
	\label{3.18}
	E^{(4)}_{\textrm{spin-2}}
	=
	\dfrac{1}{4\la}\left(
	2\dddot{h}\dot{h}
	- \ddot{h}^2
	\right)
	+ \dfrac{1}{4}
	\Big(
	\dfrac{1}{\ka^{2}}
	+\dfrac{2{k}^2}{\la}
	\Big)\dot{h}^2
	+ \dfrac{1}{4}
	\Big(
	\dfrac{{k}^{2}}{\ka^{2}}
	+\dfrac{{k}^4}{\la}
	\Big) h^2\,.
	\eeq	
	It is easy to check that the first term in this expression, with the
	highest (fourth) time derivatives, has an indefinite sign. This means,
	in the UV (i.e., at the highest frequencies) the fourth-derivative
	theory is not positively defined, as it should be expected in the
	theory with the higher-derivative ghosts. At the same time, for	$\la > 0$, both second and third terms in
(\ref{3.18}) are positively defined, indicating the positively defined energy in the low energy limit (the IR) for the spin-2 mode.
Thus, this positiveness defines the sign of $\la$, similar to how we
define the sign of the Einstein-Hilbert action in GR.

It is important to emphasize that this definition of energy scales, in which the higher time-derivative terms are associated with the UV regime, while the lower ones correspond to the IR regime, is a consequence of the violation of the standard massless dispersion relation between the wave vector and the frequency, see e.g., \cite{OUP} and references therein. In this scenario, it is possible to observe infrared behavior even when $\vec k$ takes large values.

The transition to GR in (\ref{3.18}) requires the limit
$\la \rightarrow + \infty$. Then the graviton energy for the
	Einstein-Hilbert action
	\beq
	\label{3.19} E^{(2)}_{\textrm{spin-2}}
	\,=\,
	\dfrac{1}{4\ka^2}\left(\dot{h}^2+{{k}^2}h^2\right)\, ,
	\eeq
	is positively defined for $\ka^2 > 0$, i.e., for positive Newton
	constant;
	
\textit{iii)} \ \
In general, higher-derivative gravity models exhibit an indefinite energy sign in the UV regime. In the specific case of sixth-order gravity, this indefiniteness is also present. However, if the conditions in (\ref{3conds}) are satisfied and we further assume that the wave vector $\vec{k}$ has a magnitude much larger than the two massive parameters $m_1^2$ and $m_2^2$ (i.e., $k^2\gg m_1^2$ and $k^2\gg m_2^2$), the energy expression (\ref{3.17}) reveals an unexpected behavior in the UV limit: the terms with six time derivatives are positive definite. To demonstrate this, one can replace the term $h$ by the oscillatory
solution (\ref{oscila}) into the higher-derivative term in the equation
(\ref{3.17}). Then,
\beq
\label{3.21}
E^{(6)}_{\textrm{spin-2}}
\bigg|_{ 
			\textrm{UV} 
}
 \!\!=\,
\dfrac{\te_{1,C}}{4}
\Big[  
\stackrel{...}{h}^{\,2}
-\, 2h^{\textrm{(IV)}}\ddot{h}
+ 2h^{\textrm{(V)}}\dot{h}
\Big]
\,\geq\,0\,.
\eeq

In contrast, the UV energy under analogous conditions is negative, as demonstrated below
\beq
\label{3.21-b}
E^{(4)}_{\textrm{spin-2}}
\bigg|_{ 
		\textrm{UV} 
}
\!\!=\,
\dfrac{1}{4\la}
\Big(  
2\stackrel{...}{h}\dot h
-\, \ddot{h}^{2}
\Big)
\,\leq\,0\,.
\eeq
Details of the intermediate steps used to derive expressions (\ref{3.21}) and (\ref{3.21-b}) can be found in Appendix \ref{Appendix A}.

The physical explanation of the qualitative difference between
sixth- and fourth-derivative models is as follows. According to
the analysis of \cite{highderi}, the spin-2 part of the propagator
of the model with six derivatives, in the case of real mass
spectrum, has the form
\beq
\label{3.22}
	G_{2}
	\,\, \propto  \,\, \Big(
	\dfrac{A_0}{{k}^2}
	- \dfrac{A_1}{{k}^2+m_1^2}
	+ \dfrac{A_2}{{k}^2+m_2^2}
	\Big)
	\,\hat{P}^{(2)}\,.
\eeq
Here $m_2^2 > m_1^2$ and the coefficients $\,A_0$, $\,A_1\,$ and
	$\,A_2\,$ are positive constants. This means, there is a massless
	graviton with $A_0$ (can be always set to unity by rescaling
	the quantum metric), then the ghost massive particle with $A_1$
and the heavier massive particle with $A_2$, which is healthy.
The qualitative explanation is that,
taking only the  highest frequencies modes in (\ref{3.21}),
the masses can be ignored and we get positive energy because there
are more healthy	degrees of freedom compared to the ghost degrees
of freedom.
	
As we know, the differences between ghost and tachyonic
ghost is that a ghost
	is assumed to cause instability in the classical solution only when
	it couples to healthy fields or to the background. On the other hand,
	the tachyonic ghost is unstable in all situations. This difference
	justifies the use of the positive sign of $ \la $ in the fourth-derivatives
	in the gravity model \cite{OUP,HD-Stab,Cusin}. In the six-derivative
	gravity model, an alternative with no tachyon instabilities, is when
	$\te_{1,C}$ and $\la$ have real positive values.

	Here, we make a final observation about the positivity of energy. As discussed previously, the oscillatory solutions can arise even with negative masses when the absolute value of the wave vector is sufficiently large. In this case, for the sixth-derivative model, the constant $\vartheta_{1,C}$ may take a negative value; similarly, in the fourth-derivative model, we find $\lambda < 0$. In this scenario, the energies in (\ref{3.21}) and (\ref{3.21-b}) have opposite signs. Most importantly, in the IR limit, the energy becomes indefinite. Therefore, the most consistent choice is the conditions given in (\ref{3conds});

	\textit{iv)} \ \
It is interesting that the unique intermediate term in (\ref{3.17}) with an indefinite sign is the second one, namely $2\dddot{h}\dot{h} - \ddot{h}^2$. For the six-derivative model, this term is subdominant in both the high- and low-energy limits. Formally, the positivity of the energy is guaranteed in the IR and UV regimes, as previously described, although it is not defined in intermediate energies.
	
It is important to note that the intermediate energy term in the six-derivative model coincides with the UV energy term of the four-derivative model, corresponding to the first term on the \textit{ r.h.s.} of 	(\ref{3.18}), where negative energy values occur only when the magnitude of the wave vector is far above the Planck scale. Since this condition defines a UV regime, where the mass scales are negligible, it cannot be imposed in (\ref{3.17}) to define the energy in the intermediate regime.

\subsection{Scalar mode in the six-order gravity}
\label{sec3.2}

The similarity of the bilinear forms (\ref{1.23}) in the tensor
and scalar sectors  makes it possible to perform the
analysis in a similar way.
For the scalar mode, using the second term on the \textit{r.h.s.}
of Eq.~(\ref{2.4}) and setting
$\Psi(-\cx )=-\te_{1,R}\cx-1/\xi$, with $\te_{0,R}=1/\xi$, we get
\beq
\label{4.1}
S_{\textrm{spin-0}}^{(2)}
\,\,=\,\,
-\,\dfrac{3}{2} \int d^4x
\Big\{
\te_{1,R} \big(\cx^2 \phi\big) \cx \phi
+\dfrac{1}{\xi}
\big(\cx \phi \big)^2
-\dfrac{1}{3\ka^2} \phi\cx \phi
\Big\}\,,
\eeq
where we used the abbreviation 
$\phi={\phi}_{\mu\nu}$ for the field (\ref{2.5}).

The Lagrange function of individual mode with
fixed wave vector $\vec{k}$, has the form
\beq
\label{4.2}
L_{\textrm{spin-0 }}
&=&
-\, \dfrac{3}{2}\,\te_{1,R}
\big(\phi^{\textrm{(IV)}} + 2{k}^2 \ddot{\phi} + k^4 \phi \big)
\big( \ddot{\phi} + k^2 \phi \big)
\nonumber
\\
&&
\quad
-\, \dfrac{3}{2\xi} \big( \ddot{\phi} + k^2 \phi \big)^2
\,+\,\dfrac{1}{2\ka^2}\,\phi \big(\ddot{\phi} + k^2 \phi \big) \,,
\eeq
and its  equation of motion  is given by
\beq
\label{4.3}
&&
\phi^{\textrm{(VI)}}
+ 3{k}^{2}\phi^{\textrm{(IV)}}
+ 3{k}^{4}\ddot{\phi}
+{k}^{6}\phi
+\dfrac{1}{\xi\te_{1,R}}
\big(
\phi^{\textrm{(IV)}}
+ 2{k}^{2}\ddot{\phi}
+{k}^{4}\phi
\big)
\nonumber
\\
&&
\qquad
-\,\,  \dfrac{1}{3\ka^{2}\te_{1,R}}
\big(\ddot{\phi} +{k}^{2}\phi \big) \,=\, 0 \,,
\eeq
which can be rewritten in the form
\beq
\label{4.5}
\Big(\dfrac{\pa^{2}}{\pa t^{2}}
+ {k}^{2}\Big)\Big(\dfrac{\pa^{2}}{\pa t^{2}}
+ \om_{1}^{2}\Big)\Big(\dfrac{\pa^{2}}{\pa t^{2}}
+\om_{2}^{2}\Big)\phi\,=\,0\,,
\eeq
where 
\beq
\om^2_1 = {k}^{2}+m_1^{2}
\qquad
\mbox{and}
\qquad
\om^2_2={k}^{2}+m_2^{2} \,.
\label{mas-scalar}
\eeq
As in the tensor sector, we have two massive parameters satisfying
the conditions
\beq
\label{4.4}
m_1^{2}+m_2^{2}
\,=\,\dfrac{1}{\xi\te_{1,R}}
\qquad
\mbox{and}
\qquad
m_1^{2} \, m_2^{2}
\,=\,-\dfrac{1}{3 \ka^{2}\te_{1,R}}\,.
\eeq
Solving the above equations yields
\beq
m_1^2
\,=\,
\dfrac{1}{2\xi \,\te_{1,R}}
\left(\, 1\,\pm\,
\sqrt{1 + \frac{4\xi^2 \te_{1,R}}{3\ka^2}}\,\right)\, ,
\nn
\\
m_2^2
\,=\,
\dfrac{1}{2\xi \,\te_{1,R}}
\left(\, 1\,\mp\,
\sqrt{1 + \frac{4\xi^2 \te_{1,R}}{3\ka^2}}\,\right)\, ,
\label{3.2-7}
\eeq
where $m_1^2$ and $m_2^2$ can be either real (positive or negative) or complex.

We already know, through the previous section, that the mass spectrum
defined by (\ref{mas-scalar}) and (\ref{3.2-7}) can admit both normal
oscillatory solutions (\ref{oscila}) and tachyonic-type exponential
solutions. The stability of the spin-2 mode in the IR limit, i.e., in GR,
requires $\ka^{2}>0$. For the independent parameters $\xi$ and
$\te_{1,R}$, using (\ref{mas-scalar}) and (\ref{3.2-7}), we get the
relations
\beq
\xi < 0 \,,
\qquad
\te_{1,R}<0 \,,
\qquad
\mbox{and}
\qquad
\dfrac{4\xi^2 \te_{1,R}}{3\ka^2} < 1 \,.
\label{3conds-scalar}
\eeq
With these conditions satisfied, there are periodic solutions for any magnitude value of
$\vec k$. Indeed, there are also oscillatory solutions (\ref{oscila}) with
negative $m_1^2$ or $m_2^2$, but only for a sufficiently large absolute
value of wave vector. An
important detail is that, there may be complex conjugate poles
when the last of the inequalities (\ref{3conds-scalar}) is violated.

Using the Lagrange function (\ref{4.2}) in Eq. (\ref{energy}), we
get the energy of the individual wave,
\beq
\label{4.12}
E_{\textrm{spin-0}}
&=&
\dfrac{3}{2}\,\te_{1,R}
\big(\dddot{\phi}^{2}
- 2\phi^{\textrm{(IV)}}\ddot{\phi}
+ 2\phi^{\textrm{(V)}}\dot\phi \big)
+\dfrac{3}{2}\Big( \dfrac{1}{\xi}
+3\,\te_{1,R}\, k^{2} \Big)
\big(2\dddot{\phi}\dot{\phi} - \ddot{\phi}^2 \big)
\nonumber
\\
&&
+ \,\,\dfrac{3}{2}
\Big( \dfrac{2{k}^{2}}{\xi}
- \dfrac{1}{3\ka^{2}}
+3\te_{1,R}{k}^{4}
\Big)\dot{\phi}^{2}
+\dfrac{3}{2}
\Big(
\dfrac{{k}^{4}}{\xi}
- \dfrac{{k}^{2}}{3\ka^{2}}
+ \te_{1,R}{k}^{6}
\Big)\phi^{2} \,.
\eeq
Let us point out the differences and similarities with the tensor
mode formula (\ref{3.17}):

\textit{i)} \ \
Assuming $\te_{1,R}=0$ in the six-derivative model, one recovers the energy expression for the scalar sector of the fourth-derivative gravity model,
\beq\label{4.13}
E^{\left(4\right)}_{\textrm{spin-0}}
\,=\, \dfrac{3}{\xi}
\Big(\dddot{\phi}\dot{\phi}
-\dfrac{1}{2}\ddot{\phi}^2\Big)
+ \Big(\dfrac{3k^2}{\xi}
- \dfrac{1}{2\ka^2} \Big)\dot{\phi}^2
+ \Big( \dfrac{3{k}^{4}}{2\xi} - \dfrac{k^2}{2\ka^2}
\Big)\phi^2 \,.
\eeq	
The sign of the energy is indefinite in the IR limit if the sign and magnitude of the constant $\xi$ are unknown. The same is true for the UV limit. However, for $\xi<0$, the conditions for oscillatory solutions are satisfied. Contrary to the tensor mode, in the IR limit the energy is negatively defined, this is also valid for (\ref{4.12}). On the other hand, in the UV limit, with $k^2\gg m^2_2>m^2_1$, it is positively defined,
\beq
\label{4.13-5}
E^{\left(4\right)}_{\textrm{spin-0}}\Big|_{\textrm{UV}}
\,=\, \, \dfrac{3}{2\xi}
\big(2\dddot{\phi}\dot{\phi} - \ddot{\phi}^{2}\big)\,\,
\geq\,\,0\,.
\eeq

When we take $\xi\rightarrow\infty$, the remaining expression
for the energy is given by
\beq
\label{4.14}
E_{\textrm{spin-0}}
\,=\,
- \dfrac{1}{2\ka^2}\big(\dot{\phi}^2+{{k}^2}\phi^2\big)
\,\leq \,0 \,,
\eeq
showing that the energy of the scalar mode is negative in the GR — a well-known feature.
It is important to stress that this result refers to the quantum linearized theory after applying the Faddeev–Popov procedure. In the original classical
GR, the scalar mode is gauge-fixing dependent;

\textit{ii)} \ \
In the UV limit, the energy expression in (\ref{4.12}) is dominated by the sixth derivative of $\phi$ with respect to time and is indefinite. However, if the conditions in (\ref{3conds-scalar}) are satisfied and after taking $k^2\gg m^2_2>m^2_1$, the energy of the model becomes negatively defined,
\beq
\label{4.15}
E^{\left(6\right)}_{\textrm{spin-0}}\Big|_{\textrm{UV}}
\,=\,\,\, \dfrac{3}{2}\te_{1,R}
\big( \dddot{\phi}^{2}
-2\phi^{\textrm{(IV)}}\ddot{\phi}
+2\phi^{\textrm{(V)}}\dot\phi
\big)\,\, \leq\,\,  0 \,.
\eeq
The qualitative interpretation of the results in items \textit{i} and \textit{ii}, can be easily explained using the logic of
\cite{highderi}, as we did in the previous section.
In (\ref{4.14}) the lightest scalar mode is a ghost, while the massive particle in  (\ref{4.13-5}) is healthy. For the case of the previous equation, there are more ghost degrees of freedom than healthy ones.  
The intermediate calculations in this subsection are provided in Appendix \ref{Appendix A}


\section{Positivity conditions in eight-order gravity}
\label{sec4}

We can write the spin-2 and spin-0 tensor actions of the eight-derivative gravity model using the \textit{r.h.s.} of Eq.~(\ref{2.4}) and the following settings
\beq
\Phi(-\cx)=
-\te_{2,C}\cx^2-\te_{1,C}\cx-\frac{1}{\la}
\qquad
\textrm{and}
\qquad
\Psi(-\cx)=
-\te_{2,R}\cx^2-\te_{1,R}\cx-\frac{1}{\xi} \,.
\eeq
After a straightforward algebraic manipulation, the actions for the  tensor mode $h$ and scalar mode $\phi$ become
\beq
\label{5.2}
&&
S_{\textrm{spin-2}}^{\left(2\right)}
\,=\, - \dfrac{1}{4}\int d^4x
\Big\{
\te_{2,C}\left(\cx^{2}h\right)^{2}
+ \te_{1,C}\left(\cx^{2}h\right)\cx h
+ \dfrac{1}{\la}\left(\cx h\right)^{2}+\dfrac{1}{\ka^{2}}h\cx h\Big\}\,,
\\
\label{5.2-S}
&&
S_{\textrm{spin-0}}^{\left(2\right)}
\,=\,
- \dfrac{3}{2}\int d^{4}x
\Big\{\te_{2,R}\,\left(\cx^{2}\phi\right)^{2}
+ \te_{1,R}\,\left(\cx^{2}\phi\right)\cx\phi
+ \dfrac{1}{\xi}\left(\cx\phi\right)^2
-  \dfrac{1}{3\ka^2}\,\phi\cx\phi\Big\}\,.
\qquad
\eeq

The corresponding Lagrangians for individual plane-wave modes have the
\beq
L_{\textrm{spin-2}}
&=&
-\,\dfrac{1}{4}\te_{2,C}
\Big(
h^{\textrm{(IV)}}
+ 2{k}^{2}\ddot{h}
+ {k}^{4}h \Big)^2 - \dfrac{1}{4}\te_{1,C}
\Big(h^{\textrm{(IV)}} + 2{k}^{2}\ddot{h}
+{k}^{4}h \Big) \Big(\ddot{h} + k^2 h \Big)
\nonumber
\\
&&
\quad
-\, \dfrac{1}{4\la} \Big(\ddot{h} +k^2h \Big)^2
- \dfrac{1}{4\ka^2}h
\Big( \ddot{h} + k^2 h \Big) \, ,
\label{eq:3.53}
\\
L_{\textrm{spin-0}}
&=&
-\, \dfrac{3}{2}\te_{2,R}\Big(\phi^{\textrm{(IV)}}
+ 2{k}^{2}\ddot{\phi}
+ {k}^{4}\phi\Big)^{2}
- \dfrac{3}{2}\te_{1,R}\Big(\phi^{\textrm{(IV)}}
+ 2{k}^{2}\ddot{\phi}
+ {k}^{4}\phi\Big)\Big(\ddot{\phi}+{k}^{2}\phi\Big)
\nonumber
\\
&&
\quad
-\, \dfrac{3}{2\xi}\Big(\ddot{\phi}+ k^2\phi\Big)^2
+ \dfrac{1}{2\ka^2}\phi\Big(\ddot{\phi} + k^2\phi\Big) \, .
\label{eq:3.53-S}
\eeq

The equations for both (\ref{eq:3.53}) and (\ref{eq:3.53-S})
 can be formulated for $\Phi = (h,\phi)$ as
\beq
\Big(\dfrac{\pa^2}{\pa t^2}+k^2\Big)
\Big(\dfrac{\pa^2}{\pa t^2}+\om^2_1\Big)
\Big(\dfrac{\pa^2}{\pa t^2}+\om^2_2\Big)
\Big(\dfrac{\pa^2}{\pa t^2}+\om^2_3\Big)\Phi
\,\,=\,\, 0 \,,
\eeq
where  $\om^2_{1,2,3}$ can represent the frequencies of the spin-2 or spin-0 modes, related to the mass parameters by
\beq \label{om123}
\om^2_1=k^2+m_1^2\,,
\quad
\om^2_2=k^2+m_2^2\,,
\quad
\textrm{and}
\quad
\om^2_3=k^2+m_3^2\,.
\eeq

In the tensor sector, these masses obey the relations 
\beq
\label{Cond-8T}
&&
m_1^2+m_2^2+m_3^2
\,=\, \dfrac{\te_{1,C}}{\te_{2,C}}\,,
\nonumber \\ &&
m_1^2\,m_2^2+m_1^2\,m_3^2+m_2^2\,m_3^2
\,=\,  \dfrac{1}{\la\te_{2,C}}\,,
\nonumber \\ &&
m_1^2\,m_2^2\,m_3^2
\,=\, \dfrac{1}{\ka^2\te_{2,C}}\,.
\eeq
From this system of equations, we derive the following cubic polynomial valid for each mass squared $\al \equiv m^2_i$ (with $i=1,2,3$),
\beq \label{poly-3}
	\al^3 - \dfrac{\vartheta_{1,C}}{\vartheta_{2,C}}\,\al^2
	+ \dfrac{1}{\la \vartheta_{2,C}}\, \al
	- \dfrac{1}{\ka^2 \vartheta_{2,C}}
	=
	0 \, .
	\nn
\eeq 
 The roots of this polynomial can be either real or complex, depending on the values and relations among the constants $\vartheta_{2,C}$, $\vartheta_{1,C}$ and $1/\lambda$. According to Descartes’ rule of signs (see \cite{Cheri (20)} and references therein), one way to obtain the existence of positive roots requires that these constants also be positive.

The condition that the constants are positive is consistent with the first two conditions in (\ref{3conds}) for the sixth-derivative model. Furthermore, this reasoning can be extended to the scalar mode of the theory, where the corresponding constants must be negative, as will be demonstrated later. It is also important to emphasize that this logic applies equally to models with more than eight derivatives.

The energy of individual wave with the
momentum $\vec{k}$, in the tensor sector,  is given by
\beq \label{5.4}
E^{\left(8\right)}_{\textrm{spin-2}}
&=&
\dfrac{1}{4}\, \te_{2,C}
\left[
 2h^{\left(\textrm{VII}\right)}\dot{h}
-\left(h^{\textrm{(IV)}}\right)^{2}
+ 2h^{\textrm{(V)}}\dddot{h}
- 2h^{\textrm{(VI)}}\ddot{h}
\right]
\nonumber
\\
&& +\,
\dfrac{1}{4}\left(\te_{1,C}+4\te_{2,C}k^{2}\right)\left[
\dddot{h}^{2}-2h^{\textrm{(IV)}}\ddot{h}
+ 2
h^{\textrm{(V)}}\dot{h}\right]
\nonumber
\\
&&
+ \, \dfrac{1}{4} \Big(\dfrac{1}{\la}
+ 3\te_{1,C}{k}^{2} + 6\te_{2,C}{k}^{4}\Big)
\big(2\dddot{h}\dot{h} - \ddot{h}^2\big)
\nonumber
\\
&&
+ \, \dfrac{1}{4}\Big(\dfrac{1}{\ka^{2}}
+ \dfrac{2{k}^{2}}{\la}
+ {3\te_{1,C}{k}^{4}}
+4\te_{2,C}{k}^{6}\Big)\dot{h}^{2}
\nonumber
\\
&&
+ \, \dfrac{1}{4}\Big(\dfrac{{k}^{2}}{\ka^{2}}
+ \dfrac{{k}^{4}}{\la}
+ {\te_{1,C}{k}^{6}}
+ {\te_{2,C}{k}^{8}}\Big)h^2\,.
\eeq
Taking $\te_{2,C}=0$, we recover the energy expression of the six-derivative gravity model, as discussed in the previous section. In the opposite case, where $\te_{2,C}\neq 0$, the energy sign is, in general, indefined.
Assuming the constants are such that the massive parameters are real and positive, the energy becomes negative in the UV limit, where $k^2$ is much greater than all three
massive parameters,
\beq
E^{\left(8\right)}_{\textrm{spin-2}} \Big|_{\textrm{UV}}
&=&
\dfrac{1}{4}\te_{2,C}\left[
2h^{\textrm{(V)}}\dddot{h}
- \left(h^{\textrm{(IV)}}\right)^{2}
-2h^{\textrm{(VI)}}\ddot{h}
+ 2h^{\left(\textrm{VII}\right)}
\dot{h}\right] \,\, \leq\,\,0\,.
\eeq
As expected, in the IR regime the energy is positive definite for positive constants $\te_{2,C}$, $\te_{1,C}$ and $1/\la$.

In the scalar sector, the three massive parameters in Eq.~(\ref{om123}) satisfy the following conditions
\beq
\label{Cond-8S}
&&
m_1^2+m_2^2+m_3^2
\,=\,
\dfrac{\te_{1,R}}{\te_{2,R}}\,, \nonumber
\\
&&
m_1^2\,m_2^2+m_1^2\,m_3^2+m_2^2\,m_3^2
\,=\,
\dfrac{1}{\xi\te_{2,R}}\,, \nonumber
\\
&&
m_1^2\,m_2^2\,m_3^2
\,=\,
 - \dfrac{1}{3\ka^2\te_{2,R}}\,.
\eeq
Note that the structures of system of equations (\ref{Cond-8T}) and
(\ref{Cond-8S}) are similar . The differences are in the constants
$\te_{N,C}$ and $\te_{N,R}$ with $N=0,1,2 $, and in the
Einstein-Hilbert. One can perform the following changes to
transform (\ref{Cond-8T}) to (\ref{Cond-8S}), or vice-versa,
\beq \label{Trans}
\te_{N,C} \,\longleftrightarrow\, \te_{N,R}\,,
\qquad  \textrm{and} \qquad
\dfrac{1}{\ka^2}\,\longleftrightarrow\,-\dfrac{1}{3\ka^2}\,.
\eeq
 To ensure positive \textit{r.h.s.} in 
 (\ref{Cond-8S}), like in (\ref{Cond-8T}), we must assume that
 $\te_{N,R} < 0$. We will see later on, that this makes the energy
 of the scalar mode to be negatively defined in the IR limit.
The expression for the energy of the scalar mode is
\beq
E^{\left(8\right)}_{\textrm{spin-0}}
&=&
\dfrac{3}{2}\,\te_{2,R}
\left[ 2\phi^{\textrm{(VII)}}\dot{\phi}
- \left(\phi^{\textrm{(IV)}}\right)^{2}
+ 2\phi^{\textrm{(V)}}\dddot{\phi}
- 2\phi^{\textrm{(VI)}}\ddot{\phi}
\right]
\nonumber
\\
&&
+\,\dfrac{3}{2}\left(
\te_{1,R}+4\te_{2,R}{k}^{2}\right)
\left[\dddot{\phi}^{2}
-2\phi^{\textrm{(IV)}}\ddot{\phi}
+2\phi^{\textrm{(V)}}\dot{\phi}\right]
\nonumber
\\
&&
+\, \dfrac{3}{2}\Big(\dfrac{1}{\xi}+3\te_{1,R}{k}^{2}
+6\te_{2,R}{k}^{4}\Big)
\Big( 2\dddot{\phi}\dot{\phi} - \ddot{\phi}^{2} \Big)
\nonumber
\\
&&
+\,\dfrac{3}{2}\Big(
4\te_{2,R}{k}^{6}
+ {3\te_{1,R}{k}^{4}}
+\dfrac{2{k}^{2}}{\xi}
-\dfrac{1}{3\ka^{2}}
\Big)\dot{\phi}^{2}
\nonumber
\\
&&
+ \dfrac{3}{2}\Big(
{\te_{1,R} k^6}
+ \dfrac{k^{4}}{\xi}
- \dfrac{{k}^2}{3\ka^2}
+ {\te_{2,R}{k}^{8}}
\Big)\phi^2\,.
\eeq
To analyze the energy behavior in the UV regime, we assume that the massive parameters are real, positive, and much smaller than the wave vector $\vec{k}$. Under this assumption, the leading contribution to the energy becomes
\beq
E^{\left(8\right)}_{\textrm{spin-0}} \Big|_{\textrm{UV}}
\,=\,\,
\dfrac{3}{2}\te_{2,R}
\left[
2\phi^{\textrm{(VII)}}\dot{\phi}
- \left(\phi^{\textrm{(IV)}}\right)^{2}
+ 2\phi^{\textrm{(V)}}\dddot{\phi}
- 2\phi^{\textrm{(VI)}}\ddot{\phi}
\right]\,\, \geq \,\,0\,,
\eeq
of course, if $\te_{2,R}$ is negatively defined, as previously discussed.

In the IR limit, the energy determined by the lowest-order derivative term is negatively
defined for $\te_{N,R}<0$,  similar to the fourth-order gravity
in the same regime.

For a comprehensive derivation of the results discussed in this chapter, see Appendix~\ref{Appendix A}.

\section{Beyond the eight-derivative model}
\label{sec5}

Thus far, we have investigated the conditions for the positivity of energy and stable solutions for the fourth, sixth, and eighth-derivative models in both the tensor and scalar modes. Through these studies, a standard behavior has been implicitly demonstrated in relation to these models, both in terms of the number of derivatives, as established in the previous sections, and in the transition between the tensor and scalar modes, as discussed earlier. Therefore, it is worthwhile to develop general equations to study the behavior of higher-derivative models in the UV and IR regimes, under the assumption that the stability conditions are satisfied.

Consider the equation of motion for a polynomial gravity model with $(2N+4)$ derivatives as $N \in \mathbb{N}$, where $\Phi = (h, \phi)$ denotes the spin-2 and spin-0 modes
\beq
\Big(\dfrac{\partial^{2}}{\partial t^{2}}+k^{2}\Big)
\Big(\dfrac{\partial^{2}}{\partial t^{2}}+\om_{1}^{2}\Big)
\,\cdots\,
\Big(\dfrac{\partial^{2}}{\partial t^{2}}+\om_{N+1}^{2}\Big)\Phi
\,=\,0\,,
\label{eq:5.1}
\eeq
where $\om_{N+1}^{2}$ are the frequencies for plane wave solutions of the tensor or
scalar mode. In both cases,
\beq
\om_{N+1}^{2}=k^{2}+m_{N+1}^{2}
\eeq
being $m_{N+1}^{2}$  the massive parameters, depending of
$\left(\vartheta_{N,C}\,,\,1/\kappa^{2}\right)$ for the tensor mode and of $\left(\vartheta_{N,R}\,,\,1/\kappa^{2}\right)$ for the scalar mode.

To avoid tachyonic ghosts, we have to assume that $m_{N+1}^2$
has real positive values, that requires 
$\te_{N,C}>0$ for tensor modes and 
$\te_{N,R}<0$ for scalar modes.
In this way, we find an oscillation solution for (\ref{eq:5.1}),
i.e.,
\beq
\Phi\left(t\right)=C_{0}\cos\left(k\,t+\phi_{0}\right)
+C_{1}\cos\left(\om_{1}\,t+\phi_{1}\right)
+\ldots
+C_{N+1}\cos\left(\om_{N+1}\,t+\phi_{N+1}\right) \,.
\label{eq:5.3}
\eeq
Observing the structure of the energy equations in the UV limit established in the previous sections, we can write the general expression for the energy in this regime
\beq
E^{\left(2N+4\right)}\Big|_{\textrm{UV}}
\,\,=\,\,- \,(-1)^{N}
\Th_{N}\bigg[
\left(\Phi^{(N+2)}\right)^{2}+2\sum_{j=1}^{m}
\left(-1\right)^{j}\Phi^{(N+j+2)}\Phi^{(N-j+2)}\bigg] \,,
\label{eq:5.4}
\eeq
with $m=N+1$. The expression (\ref{eq:5.4}) applies to both tensor and scalar 
components of the plane wave, in any polynomial QG theory.
For the tensor mode $\Phi\left(t\right)=h\left(t\right)$, we
identify 
$\Th_{N}=\frac{1}{4}\te_{N,C}$,
whereas for the scalar mode $\Phi\left(t\right)=\phi\left(t\right)$,
the identification is $\Th_{N}=\frac{3}{2}\te_{N,R}$.

Assuming that $\Phi\left(t\right)$ is given by
Eq.~(\ref{eq:5.3}), in the UV regime with $k^2 \gg m^{2}_{N+1}$,
the following general rules hold:
\\		
\textit{i)} \
For the tensor mode, the energy is negatively defined when $N$
is even, and positively defined when $N$ is odd;
\\ 		
\textit{ii)} \
For the scalar mode, the rule is opposite, as the energy being 
positively defined for even $N$ and negatively defined for odd $N$.
Let us note that this conclusion is valid only for the real spectrum.

The origin of the described general rule is the sign alternation
theorem of the generalization of relation (\ref{3.22}). The
corresponding general formula has the form \cite{highderi}
\beq
G
\propto
\dfrac{A_{0}}{k{{}^2}}
+\dfrac{A_{1}}{k^{2}+m_{1}^{2}}
+\dfrac{A_{2}}{k^{2}+m_{2}^{2}}
+ \ldots
+\dfrac{A_{N+1}}{k^{2}+m_{N+1}^{2}} \,,
\eeq
where $m_{N+1}^2$ follow 
$\,0<m_1^2  < m_{2}^{2}
<\ldots<m_{N+1}^{2}$ and the signs of $A_k$
alternate, i.e., $A_k\, A_{k+1} < 0$.

In the tensor sector, the lightest particle is the healthy graviton,
and the massive degrees of freedom, for a real spectrum, follow
the hierarchy of masses. In the UV limit  with odd $N$, this
means more healthy degrees of freedom than ghosts and, for
the same 
$\vec{k}$, guarantees positivity of the total energy.
For even $N$, there is an equal number of healthy and
ghost degrees of freedom, and the greater masses of ghosts
lead to the negative overall energy.
For the scalar degrees of freedom, the difference is that the
lightest mode has a negative energy. This provides the
aforementioned flip of signs in the cases of even and odd $N$.

One can write a general expression for the energy of the plane
wave in the IR regime,
\beq
E^{\left(2N+4\right)}\Big|_{\textrm{IR}}
\,\,=\,\,
\bigg[\sum_{j=0}^{N}\left(2+j\right)\Th_{j}k^{2j+2}
+\rho \bigg]
\dot{\Phi}^{2}
\,+\,
\Big(\sum_{j=0}^{N}\Th_{j}\,k^{2j+4}
+\rho k^{2} \Big)\Phi^2 \,.
\label{genIR}
\eeq
Similar to Eq.~(\ref{eq:5.4}), this formula applies to both tensor
and scalar modes, with the following identifications:
\\
\ \textit{i)}
For tensor modes, $\Phi\left(t\right)=h\left(t\right)$ for the field,
$\Th_{N}=\frac{1}{4}\te_{N,C}$, and $\rho=\frac{1}{4\ka^{2}}$;
\\
\textit{ii)}
For the scalar degrees of freedom,
$\Phi(t)=\phi(t)$ for the field, $\Th_{N}=\frac{3}{2}\te_{N,R}$,
and $\rho=-\frac{1}{2\ka^{2}}$.

It is easy to see that the energy in the IR limit for tensor mode
is positively defined for $\te_{N,C}>0$. Indeed, the IR positivity
in the tensor sector is expected from the universality of quantum
GR as effective theory of QG \cite{don,EffAM}, so our conclusion
confirms this important feature.
Thus, the violation of positivity may take
place only at the intermediate energies.   		
Finally, the scalar mode is negatively defined for $\te_{N,R}<0$.
Both observations are in agreement with the signs of the energy of
the corresponding modes in linearized GR.

\section{Conclusions}
\label{sec6}

The standard method to evaluate the degrees of freedom of a
higher derivative model is by introducing the set of auxiliary fields
and assuming these fields evolve independently from each other
(see, e.g., \cite{Cremin}).
This approach leads to the well-known Ostrogradsky instabilities
in classical mechanics \cite{Ostrog}, which are assumed to persist
in the more complicated field theory models, at both classical and
quantum levels \cite{Woodard-review}.
The situation may be quite different if we
deal with the six- or higher-derivative models using original
variables. The origin of this difference is that, in this case, the
massive and massless modes have to stay together, as it was
originally discussed in \cite{Hawking-Whois}. For the plane wave
solutions, there are constraints requiring all modes to have the same
wave vector and the same frequency.

Using this approach, we explored the positivity of the energy
of the plane wave solutions in the tensor and scalar, gauge-fixing
independent, sectors in the theory of superrenormalizable gravity
models with a polynomial Lagrangian. We have found a general
pattern for the energy positivity, which is typical only in these
models and is different from the ``minimal'' fourth-derivative QG.
One of the results is the requirement to have opposite sign
rules in the tensor and scalar sectors. We have found that this
feature is a consequence of the sign alternating theorem of
\cite{highderi}.

\section{Acknowledgments}	

P.R.B.R.V. is grateful to B.L. Giacchini and W.C. Silva for useful and valuable discussions. The author would also like to thank I.L. Shapiro for his comments and encouragement  in the
course of this work, and M.R. do Vale for constant support.

\section{Data Availability Statement}

No Data associated in the manuscript.
 
\appendix
		
\section*{Appendix}
		\addcontentsline{toc}{section}{Appendices}
		\renewcommand{\thesubsection}{\Alph{subsection}}

\subsection{Intermediate calculations of energy equations
in the UV regime}
\label{Appendix A}
		
This appendix provides details on the computation of energy conditions in the UV regime for higher-derivative models. The
expression for energy of the tensor  $h\left(t\right)$ and $\phi\left(t\right)$ modes depends
on the constants $\te_{N,C}$ and $\te_{N,R}$.
To streamline the presentation, we perform the calculations using a unified notation for
$\Phi=(h,\phi)$ and use generalized constant $\Th_{N}$, where 
$\Th_{N}=\frac{1}{4}\te_{N,C}$ for the tensor mode and 
$\Th_{N}=\frac{3}{2}\te_{N,R}$ for the scalar mode with $N \in \mathbb{N}$.

The energy in the UV limit for the fourth-derivative model is given by
\beq
E^{\left(4\right)}\Big|_{\textrm{UV}}=
\Te_{0}\left[
2\dddot{\Phi}\dot{\Phi}
 - \big(\ddot{\Phi}\big)^2\right]\,.
\eeq
If the term $\Phi$ is described by the oscillation solution
(\ref{eq:5.3}),  we arrive at the following result
\beq \label{Ap:2}
&&
E^{\left(4\right)}\Big|_{\textrm{UV}} \,=\,
-\,\,\Th_{0}\big\{ \big[C_{0}k^{2}\cos\left(kt+\ph_{0}\right)
+ C_{1}\om^2_1 \cos\left(\om_{1}t+\ph_{1}\right) \big]^2
\nonumber
\\
&&
\qquad
\qquad
\qquad
+ \,\,2\big[C_{0}k\sin\left(kt+\ph_{0}\right)
\,+\,C_{1}\om_{1}\sin\left(\om_{1}t+\ph_{1}\right)\big]
\nn \\
&&
\qquad
\qquad
\qquad
\,\,\,
\times
\left[C_{0}k^{3}\sin\left(kt+\ph_{0}\right)
+C_{1}{\om}_{1}^{3}\sin\left(\om_{1}t+\ph_{1}\right)\right]\Big\} \,.
\eeq
Here and below we use the notations
\beq
\om_1
=  m_{1}\sqrt{r^{2}+1}\,,
\quad
r=\frac{k}{m_1}\,,
\quad
{\nu}=\sqrt{r^2+1}\,
\quad
\textrm{and}
\quad
\tau=m_1t \,.
\label{om1r}
\eeq
When $k^2\gg m^2_1$, we consider $r\gg1$, which enables
the simplification $r^{2}+1 \longrightarrow r^2$ for the expressions
outside the trigonometric functions. After a small amount of algebra, we obtain the result
\beq \label{Ap:2b}
&&
E^{\left(4\right)}\Big|_{\textrm{UV}}  
\,=\,
-\,m_{1}^{4}r^4\,
\Th_{0}\Big\{
\big[C_0 \cos\left(r\tau+\ph_{0}\right)
+ C_1 \cos\left(\nu\tau+\varphi_{1}\right)\big]^2
\nn
\\
&&
\qquad
\qquad
+\,\,
2	\big[C_{0}\sin\left(r\tau+\ph_{0}\right)
+C_{1}\sin\left(\nu\tau+\ph_{1}\right)\big]^2\Big\} \,,
\eeq
the sign of the last expression depends on $\Th_{0}$.
In case $\Th_0<0$, the energy is positive.

For the sixth-derivative model the expression for the energy is given by
\beq
E^{\left(6\right)}\Big|_{\textrm{UV}}
\,=\,
\Th_1 \big[
 \dddot{\Phi}^2 - 2\Phi^{(\textrm{IV})}\ddot{\Phi}
 +2\Phi^{(\textrm{V})}\dot{\Phi}\big] \,,
\label{E6-ini}
\eeq
using the same approach as for the fourth-derivative model, we get
\beq
&&	
E^{\left(6\right)}\Big|_{\textrm{UV}} \,= \,
\Th_{1}\Big\{
\big[C_{0}k^{3}\cos\left(kt+\ph_{0}\right)
+ C_{1}\om_{1}^{3}\cos\left(\om_{1}t+\ph_{1}\right)
+ C_{2}\om_{2}^{3}\cos\left(\om_{2}t+\ph_{2}\right)\big]^2
\nn
\\
&&
\qquad \qquad
+\, 2\left[C_{0}k^{2}\cos\left(kt+\ph_{0}\right)
+ C_{1}\om_{1}^{2}\cos\left(\om_{1}t+\ph_{1}\right)
+ C_{2}\om_{2}^{2}\cos\left(\om_{2}t+\ph_{2}\right)\right]
\nn \\
&&
\qquad \quad\qquad
\times\left[C_{0}k^{4}\cos\left(kt+\ph_{0}\right)
+ C_{1}\om_{1}^{4}\cos\left(\om_{1}t+\ph_{1}\right)
+ C_{2}\om_{2}^{4}\cos\left(\om_{2}t+\ph_{2}\right)\right]
\nn \\
&&
\qquad \qquad
+\, 2\left[C_{0}k\sin\left(kt+\ph_{0}\right)
+ C_{1}\om_{1}\sin\left(\om_{1}t+\ph_{1}\right)
+ C_{2}\om_{2}\sin\left(\om_{2}t+\ph_{2}\right)\right]
\nn
\\
&&
\qquad \quad\qquad
\, \times\left[C_{0}k^{5}\sin\left(kt+\ph_{0}\right)
+ C_{1}\om_{1}^{5}\sin\left(\om_{1}t+\ph_{1}\right)
+ C_{2}\om_{2}^{5}\sin\left(\om_{2}t+\ph_{2}\right)\right]
\Big\} \,.
\qquad
\eeq
Introducing the notations
\beq \label{Ap:7}
\om_{1}=m_{2}\sqrt{r^{2}+\dfrac{{m}_{1}^{2}}{{m}_{2}^{2}}}
\quad
\om_{2}=m_{2}\sqrt{r^{2}+1}\,,
\quad
\textrm{and}
\quad
r=\frac{k}{M}\,,
\eeq
where $M$ represents the mass of the heaviest degree of freedom
in the higher-derivative model. In the present case, $M=m^2_2$ .
This notation facilitates a comparison between the masses $m_1$
and $m_2$ with the wave vector $\vec{k}$.

In the UV limit, we consider $k\gg m_2 > m_1$ owing to the mass
hierarchy. This is equivalent to $r \gg m_1/m_2$ and $r\gg1$. As before,
we implement the approximation
$r^2 + m_1^2/m_2^2\longrightarrow r^2$
and $r^{2}+1\longrightarrow r^{2}$, for the expressions outside
the trigonometric functions. Thus, we get
\beq 
\label{Ap:8}
&&
E^{\left(6\right)}\Big|_{\textrm{UV}}
\,=\, m_2^6\,r^6\,\Th_{1}
\Big\{
3 \big[C_{0}\cos\left(r\tau+\ph_{0}\right)
+ C_1\cos\left(\nu_{1,2}\tau+\ph_{1}\right)
+ C_2\cos\left(\nu\tau+\ph_{2}\right)\big]^2
\nn \\
&&
\qquad
\qquad
\qquad
+ \,2\big[C_{0}\sin\left(r\tau+\ph_{0}\right)
+ C_{1}\sin\left(\nu_{1,2}\tau+\varphi_{1}\right)
+ C_{2}\sin\left(\nu\tau+\varphi_{2}\right)\Big]^2
\Big\}\,,
\qquad
\quad
\eeq
with
\beq
\nu=\sqrt{r^2+1}\,
\quad
\nu^2_{1,2} = r^2+\dfrac{m^2_1}{m^2_2}\,
\quad
\textrm{and}
\quad
\tau = m_2\, t \,.
\nn
\eeq
Similar to the previous
example, the sign of the energy depends only on $\Th_1$.
The energy is defined positively if $\Th_{1}>0$.

Finally, for the eight-order gravity, the energy is given by
\beq
 E^{\left(8\right)}\Big|_{\textrm{UV}}
\,=\,
\Th_2 \left[\left(\Phi^{(\textrm{IV})}\right)^2
+ 2\Phi^{(\textrm{V})}\ddot{\Phi}
- 2\Phi^{(\textrm{VI})}\ddot{\Phi}
+ 2\Phi^{(\textrm{VII})}\dot{\Phi}\right]\,,
\label{E8ini}
\eeq
and applying the same approach to $\Phi$ as before, we get
\beq
&&
E^{\left(8\right)}\Big|_{\textrm{UV}}  
\,=\, 
-\,\,\Th_2
\Big\{
\big[C_{0}k^{4}\cos\left(kt+\ph_{0}\right)
+ C_{1}\om_{1}^{4}\cos\left(\om_{1}t+\ph_{1}\right) + C_{2}\om_{2}^{4}\cos\left(\om_{2}t+\ph_{2}\right)
\nn
\\
&&
\qquad \qquad
+ \,
C_{3}\om_{3}^{4}\cos\left(\om_{3}t+\ph_{3}\right)\big]^2
+ 2\big[C_{0}k^{2}\cos\left(kt+\ph_{0}\right)
+ C_{1}\om^{2}_{1}\cos\left(\om_{1}t+\ph_{1}\right)
\nn \\
&&
\qquad \qquad
+ \,
C_{2}\om_{2}^{2}\cos\left(\om_{2}t+\ph_{2}\right)
+ C_{3}\om_{3}^{2}\cos\left(\om_{3}t+\ph_{3}\right)\big]
\big[C_{0}k^{6}\cos\left(kt+\ph_{0}\right)
\nn
\\
&&
\qquad \qquad
+ \,
C_{1}\om_{1}^{6}\cos\left(\om_{1}t+\ph_{1}\right)
+ C_{2}\om_{2}^{6}\cos\left(\om_{2}t+\ph_{2}\right)
+ C_{3}\om_{3}^{6}\cos\left(\om_{3}t+\ph_{3}\right)\big]
\nn \\
&&
\qquad \qquad
+ \,
2\big[ C_{0}k\sin\left(kt+\ph_{0}\right)
+C_{1}\om_{1}\sin\left(\om_{1}t+\ph_{1}\right)
+C_{2}\om_{2}\sin\left(\om_{2}t+\ph_{2}\right)
\nn\\
&&
\qquad \qquad
+ \,
C_{3}\om_{3}\sin\left(\om_{3}t+\ph_{3}\right)\big]
\big[C_{0}k^{7}\sin\left(kt+\ph_{0}\right)
+ C_{1}\om_{1}^{7}\sin\left(\om_{1}t+\ph_{1}\right)
\nn
\\
&&
\qquad \qquad
+ \,
C_{2}\om_{2}^{7}\sin\left(\om_{2}t+\ph_{2}\right)
+ C_{3}\om_{3}^{7}\sin\left(\om_{3}t+\ph_{3}\right)\big]
+ 2\big[C_{0}k^{5}\sin\left(kt+\ph_{0}\right)
\nn
\\
&&
\qquad \qquad
+ \,
C_{1}\om_{1}^{5}\sin\left(\om_{1}t+\ph_{1}\right)
+ C_{2}\om_{2}^{5}\sin\left(\om_{2}t+\ph_{2}\right)
+ C_{3}\om_{3}^{5}\sin\left(\om_{3}t+\varphi_{3}\right)\big]
\nn
\\
&&
\qquad \qquad
\,
\times \Big[C_{0}k^{3}\sin\left(kt+\ph_{0}\right)
+ C_{1}\om_{1}^{3}\sin\left(\om_{1}t+\ph_{1}\right) + C_{2}\om_{2}^{3}\sin\left(\om_{2}t+\ph_{2}\right)
\nn
\\
&&
\qquad \qquad
+ \,
C_{3}\om_{3}^{3}\sin\left(\om_{3}t+\ph_{3}\right)\big]
\Big\} \,.
\eeq
When the wave vector is much larger than the masses, we can make the substitutions
\beq
r^2 + \frac{m_2^2}{m_3^2}\longrightarrow r^2 \,,
\qquad
r^2 + \frac{m_1^2}{m_3^2}\longrightarrow r^2 \,,
\qquad
\mbox{and}
\quad
r^2+1 \longrightarrow r^2 \,,
\eeq
assuming that $\,r=k/m_3 \gg 1$. In this way, the previous equation reduces to
\beq 
\label{Ap:11}
&&
E^{\left(8\right)}\Big|_{\textrm{UV}}
\,=\,
-\,\,
m_{3}^{8}\,r^{8}\,\Th_{2}
\Big\{ 3 \big[C_{0}\cos\left(r\tau+\ph_{0}\right)
+ C_{1}\cos\left(\nu_{1,3}\tau+\ph_{1}\right)
+ C_{2}\cos\left(\nu_{2,3}\tau+\ph_{2}\right)
\nn \\
&&
\qquad
\qquad
\qquad
+ \,
C_{3}\cos\left(\nu\tau+\ph_{3}\right)\big]^2
+ 4\big[C_{0}\sin\left(r\tau+\ph_{0}\right)
+ C_{1}\sin\left(\nu_{1,3}\tau+\ph_{1}\right)
\nn
\\
&&
\qquad
\qquad
\qquad
+ \,
C_{2}\sin\left(\nu_{2,3}\tau+\ph_{2}\right)
+C_{3}\sin\left(\nu\tau+\ph_{3}\right)\big]^2
\Big\} \,,
\eeq
with
\beq 
\label{A:12}
\nu=\sqrt{r^2+1} \,,
\quad
\nu^2_{1,3} = r^2+\dfrac{{m}^{2}_{1}}{{m}^{2}_{3}} \,,
\quad
\nu^2_{2,3} = r^2+\dfrac{{m}^{2}_{2}}{{m}^{2}_{3}} \,,
\quad	
\textrm{and}
\quad
\tau = m_3\, t.
\eeq
Thus, we find that if $\Theta_2>0$, the energy is 
negative, whereas if $\Theta_2<0$, the energy is positively 
defined. 


\end{document}